\documentclass[twocolumn ,showpacs,preprintnumbers,  
               superscriptaddress]{revtex4-1}
\usepackage{graphicx, fancybox}
\usepackage{amsmath,amssymb}

\makeatletter
\newsavebox{\@brx}
\newcommand{\llangle}[1][]{\savebox{\@brx}{\(\m@th{#1\langle}\)}%
  \mathopen{\copy\@brx\mkern2mu\kern-0.9\wd\@brx\usebox{\@brx}}}
\newcommand{\rrangle}[1][]{\savebox{\@brx}{\(\m@th{#1\rangle}\)}%
  \mathclose{\copy\@brx\mkern2mu\kern-0.9\wd\@brx\usebox{\@brx}}}
\makeatother

\begin{document}

\title{
Efficient formulas for efficiency correction of cumulants}

\author{Masakiyo Kitazawa}
\email{kitazawa@phys.sci.osaka-u.ac.jp}
\affiliation{
Department of Physics, Osaka University, Toyonaka, Osaka 560-0043, Japan}

\begin{abstract}

We derive formulas which connect cumulants of particle numbers 
observed with efficiency losses with the original ones 
based on the binomial model.
These formulas can describe the case with multiple 
efficiencies in a compact form.
Compared with the presently suggested ones based on factorial
moments, these formulas would drastically reduce 
the numerical cost for efficiency corrections 
when the order of the cumulant and 
the number of different efficiencies are large.
The efficiency correction with realistic $p_T$-dependent efficiency 
would be carried out with the aid of these formulas.

\end{abstract} 
\date{\today}
\maketitle

\section{Introduction}
\label{sec:intro}

Fluctuations, especially those of conserved charges, are promising 
observables for the analysis of primordial thermodynamics 
in relativistic heavy-ion collisions \cite{Stephanov:1999zu,
Asakawa:2000wh,Jeon:2000wg}.
In particular, non-Gaussian fluctuations characterized by 
higher order cumulants acquire much attention recently
\cite{Ejiri:2005wq,Stephanov:2008qz,Asakawa:2009aj}.
Active studies have been carried out experimentally
\cite{STAR,ALICE,Adamczyk:2013dal,Adamczyk:2014fia,Adare:2015aqk,Luo:2015ewa}
and by lattice QCD numerical simulations
\cite{Ding:2015ona,Borsanyi:2015axp}.
See for reviews \cite{Koch:2008ia,Asakawa:2015ybt} and
latest progress \cite{Luo:2015doi,Nahrgang:2016ayr}.

Experimentally, the fluctuations are measured 
by the event-by-event analysis.
In this analysis,
the number of particles arriving at some range
of detectors are counted in each event.
The event-by-event distribution characterized by the histogram
of the particle number is called the event-by-event fluctuations.
The cumulants of the particle number are constructed from 
the histogram.

In real experiments, however, particle numbers in each 
event cannot be measured accurately, because the detectors
can measure particles with some probability called efficiency
which is less than unity \cite{Abelev:2008ab}.
Due to the imperfect measurement, the event-by-event histogram 
and accordingly the cumulants constructed from the histogram 
are modified in a nontrivial way.

The effect of the efficiency on cumulants can be understood if it is
assumed that the efficiency for individual particles are uncorrelated, 
i.e. the probability to observe different particles in an event 
is not correlated with one another.
In this case, the probability distribution 
of experimentally-observed particle numbers can be
related to the original one without efficiency loss 
using binomial distribution function \cite{Kitazawa:2012at}.
In this paper, we call this relation the binomial model.
The binomial model enables us to relate the cumulants of 
the original and observed particle number distributions.
It has been recognized that 
the imperfect efficiency can significantly modify the values of 
the cumulants especially for higher order ones 
\cite{Kitazawa:2012at,Bzdak:2012ab,Luo:2014rea}.
The efficiency correction in experimental analyses thus is 
an important procedure to obtain the relevant values of the 
cumulants.
The binomial model is then extended to the case with 
multiple values of local efficiencies for different particle
species and phase spaces \cite{Bzdak:2013pha,Luo:2014rea}.

In Refs.~\cite{Bzdak:2013pha,Luo:2014rea}, the formulas for 
the efficiency correction with multiple efficiencies 
are obtained using factorial moments.
These formulas consist of $M^m$ factorial moments 
for $m$-th order cumulants with $M$ different efficiencies.
In practical analyses for efficiency correction, therefore, 
the numerical cost for the efficiency correction increases rapidly 
as $M$ becomes larger.
The number of efficiencies $M$ thus is limited to small values
in the present experimental analyses \cite{private}.

In the present study, we derive a set of formulas 
which relate the cumulants of original and observed 
distribution functions with multiple efficiencies.
They are derived by a straightforward extension 
of the method used in Ref.~\cite{Kitazawa:2012at}.
In these formulas, the original cumulants before the efficiency
loss are represented by the mixed cumulants of observed 
particles.
The number of the cumulants in these relations does not depend on $M$.
They thus would enable one to carry out the efficiency correction 
more effectively for large $M$ and higher order cumulants.
Even the realistic $p_T$-dependent efficiency \cite{Abelev:2008ab} 
would be treated with these formulas.

This paper is organized as follows.
In Sec.~\ref{sec:main}, we first present our main results,
i.e. the formulas to relate the original and observed cumulants.
Their derivation are then discussed in later sections.
Section~\ref{sec:basic} is devoted to reviews on 
the definition and properties of cumulants.
We then present the derivation for single-variable 
distribution functions in Sec.~\ref{sec:single}, as a simple
illustration of the full derivation addressed in Sec.~\ref{sec:mult}.
Section~\ref{sec:summary} is devoted to discussions
and a short summary.

\section{Main result}
\label{sec:main}

In this section, we clarify the problem considered in this paper 
and show the answer, which is the main result of this paper.

\subsection{Problem}
\label{sec:main1}

We consider a probability distribution function
\begin{align}
P(N_1,N_2,\cdots,N_M) = P(\vec{N}),
\label{eq:P}
\end{align}
for $M$ integer stochastic variables, $N_1,~N_2,\cdots,N_M$,
with $\vec{N}=(N_1,N_2,\cdots,N_M)$ being the vectorical 
representation of the stochastic variables.
In the following we call $N_i$ the number
of $i$-th particle in an event.
The purpose of the present study is to obtain the cumulants of
a linear combination of $N_i$,
\begin{align}
Q = \sum_{i=1}^M a_i N_i ,
\label{eq:Q}
\end{align}
with $a_i$ being numerical numbers.

If $P(\vec{N})$ is obtained directly in an experiment 
as an event-by-event histogram, the cumulants of $Q$ can, 
of course, be constructed straightforwardly from the histogram.
In real experiments, however, the detectors 
miss the measurement of particles with some efficiency.
The number of the $i$-th particles observed by the detector, 
$n_i$, is thus different from and smaller than $N_i$.
The probability distribution function 
of observed particle numbers obtained by the imperfect experiment
\begin{align}
\tilde{P}(n_1,n_2,\cdots,n_M) = \tilde{P}(\vec{n}),
\label{eq:tildeP}
\end{align}
differs from Eq.~(\ref{eq:P}).

If the efficiencies for the observation of individual particles 
are assumed to be independent with one another,
one can relate $\tilde{P}(\vec{n})$ and $P(\vec{N})$ in a simple form.
In this case, the probability distribution of $n_i$ for a fixed $N_i$ 
obeys the binomial distribution function 
\begin{align}
B_{p,N}(n) = \frac{N!}{n!(N-n)!} p^n (1-p)^n ,
\label{eq:binomial}
\end{align}
with $N=N_i$ and $p=\epsilon_i$ being the efficiency of 
the $i$-th particle.
Therefore, $P(\vec{N})$ and $\tilde{P}(\vec{n})$ are related 
with each other as \cite{Kitazawa:2012at,Bzdak:2013pha,Luo:2014rea}
\begin{align}
\tilde{P}(\vec{n}) = \sum_{N_1,\cdots,N_M}
P(\vec{N}) \bigg( \prod_i B_{\epsilon_i,N_i}(n_i)\bigg).
\label{eq:tildeP=PB}
\end{align}
In this paper we refer to Eq.~(\ref{eq:tildeP=PB}) 
as the binomial model.
The purpose of the present study is to represent the cumulants of 
$Q$ using those of $\tilde{P}(\vec{n})$ in the binomial model 
in a compact form.

\subsection{Answer}
\label{sec:main2}

The cumulants of $Q$ up to fourth order are given by
\begin{align}
\langle Q \rangle_{\rm c} 
=& \llangle q_{(1)} \rrangle_{\rm c} ,
\label{eq:Q1}
\\
\langle Q^2 \rangle_{\rm c} 
=& \llangle q_{(1)}^2 \rrangle_{\rm c} - \llangle q_{(2)} \rrangle_{\rm c} ,
\label{eq:Q2}
\\
\langle Q^3 \rangle_{\rm c} 
=& \llangle q_{(1)}^3 \rrangle_{\rm c} - 3 \llangle q_{(2)} q_{(1)} \rrangle_{\rm c} 
+ \llangle 3 q_{(2,1|2)} - q_{(3)} \rrangle_{\rm c} ,
\label{eq:Q3}
\\
\langle Q^4 \rangle_{\rm c} 
=& \llangle q_{(1)}^4 \rrangle_{\rm c} - 6 \llangle q_{(2)} q_{(1)}^2 \rrangle_{\rm c} 
+ 12 \llangle q_{(2,1|2)} q_{(1)} \rrangle_{\rm c} 
\nonumber \\
&
+ 6 \llangle q_{(1,1|2)} q_{(2)} \rrangle_{\rm c} 
-4 \llangle q_{(3)} q_{(1)} \rrangle_{\rm c} - 3 \llangle q_{(2)}^2 \rrangle_{\rm c} 
\nonumber \\
&
+ \llangle -18 q_{(2,1,1|2,2)} + 6 q_{(2,1,1|3)} + 4 q_{(3,1|2)} 
\nonumber \\
&
+ 3 q_{(2,2|2)} - q_{(4)} \rrangle_{\rm c} ,
\label{eq:Q4}
\end{align}
where the cumulants $\langle\cdot\rangle_{\rm c}$ and 
$\llangle\cdot\rrangle_{\rm c}$ are taken 
for $P(\vec{N})$ and $\tilde{P}(\vec{n})$, respectively.
$q_{(\cdots)}$ are linear combinations of $n_i$
defined by
\begin{align}
q_{(s)} 
=& \sum_{i=1}^M c_{(s)}^{(i)} n_i,
\label{eq:q(s)}
\\
q_{(s_1,\cdots,s_j|t_1,\cdots,t_k)} 
=& \sum_{i=1}^M c_{(s_1,\cdots,s_j|t_1,\cdots,t_k)}^{(i)} n_i.
\label{eq:q(sstt)}
\end{align}
The coefficients $c_{(\cdots)}^{(i)}$ are numerical numbers 
which depend on $a_i$ and $\epsilon_i$ defined by
\begin{align}
c_{(s)}^{(i)} =& \tilde{a}_i^s \tilde\xi_s^{(i)} ,
\\
c_{(s_1,\cdots,s_j|t_1,\cdots,t_k)}^{(i)}
=& \tilde{a}_i^{s_1+\cdots+s_j} \tilde\xi_{s_1}^{(i)} \cdots \tilde\xi_{s_j}^{(i)} 
\tilde\xi_{t_1}^{(i)} \cdots \tilde\xi_{t_k}^{(i)} ,
\end{align}
with 
$\tilde{a}_i = a_i/\epsilon_i$,
$\tilde{\xi}_m^{(i)}=\xi_m^{(i)}/\epsilon_i$, and 
$\xi_m^{(i)}=\xi_m(\epsilon_i)$ being coefficients of 
the binomial cumulants with probability $\epsilon_i$;
explicit forms of $\xi_m$ up to sixth order are given 
in Eqs.~(\ref{eq:xi1}) -- (\ref{eq:xi6}).

When one applies Eqs.~(\ref{eq:Q1}) -- (\ref{eq:Q4}) to the 
efficiency correction in event-by-event analyses, 
the required procedure is as follows:
\begin{enumerate}
\item
Calculate $q_{(\cdots)}$ in each event from $n_i$ 
observed experimentally in the event.
\item
Using the event-by-event distribution of $q_{(\cdots)}$,
obtain the (mixed) cumulants of $q_{(\cdots)}$ 
which appear in Eqs.~(\ref{eq:Q1}) -- (\ref{eq:Q4}).
\end{enumerate}
Remark that the numerical cost for the latter procedure 
does not depend on $M$.

In Ref.~\cite{Kitazawa:2012at}, 
the relation of the cumulants for the ``net-particle number'' 
with common efficiencies for particles and anti-particles, 
respectively, are derived.
This result is reproduced by substituting $M=2$, $a_1=1$ and $a_2=-1$ 
in Eqs.~(\ref{eq:Q1}) -- (\ref{eq:Q4}).

In the rest of this paper,
we derive Eqs.~(\ref{eq:Q1}) -- (\ref{eq:Q4}).

\section{Cumulants}
\label{sec:basic}

In this section we first summarize properties of cumulants 
required for the derivation of Eqs.~(\ref{eq:Q1}) -- (\ref{eq:Q4}).

\subsection{Definition}
\label{sec:cumulants}

The cumulants are defined from the cumulant generating function.
For the probability distribution function Eq.~(\ref{eq:P}),
the generating function is defined by
\begin{align}
K(\theta_1,\cdots,\theta_M)
= \ln \bigg[ \sum_{N_1,\cdots,N_M}
P(\vec{N}) \exp( \sum_{i=1}^M N_i \theta_i ) \bigg].
\label{eq:K}
\end{align}
The cumulants are then defined by the derivatives of 
Eq.~(\ref{eq:K}) as 
\begin{align}
\langle N_m^s \rangle_{\rm c}
= \frac{\partial^s}{\partial \theta_m^s} K(\vec{\theta})|_{\vec{\theta}=0}
\equiv \partial_m^s K(\vec{0}),
\label{eq:cumulant}
\end{align}
and 
\begin{align}
\langle N_{m_1}^{s_1} \cdots N_{m_j}^{s_j}  \rangle_{\rm c}
= \partial_{m_1}^{s_1} \cdots \partial_{m_j}^{s_j} K(\vec{0}).
\end{align}
From these definitions, one immediately obtains 
the ``distributive property'' of the cumulants such as 
\begin{align}
\lefteqn{ \langle (a_1 N_1 + a_2 N_2 )^2 \rangle_{\rm c} }
\nonumber \\
&= a_1^2 \langle N_1^2 \rangle_{\rm c}
+ 2 a_1 a_2 \langle N_1 N_2 \rangle_{\rm c}
+ a_2^2 \langle N_2^2 \rangle_{\rm c} .
\end{align}
Other properties of cumulants, such as their relation
with moments, are found in Ref.~\cite{Asakawa:2015ybt}.

\subsection{Cumulant expansion}
\label{sec:exp}

Consider a (non-homogeneous) linear function of $\vec{N}$, 
\begin{align}
L(\vec{N}) = d_0 + \sum_i d_{i=1}^M N_i,
\end{align}
with numerical numbers $d_i$.
The cumulants of $L(\vec{N})$ are given by
\begin{align}
\langle (L(\vec{N}))^m \rangle_{\rm c}
= \frac{\partial^m}{\partial \bar\theta^m} K_L(\bar\theta) |_{\bar\theta=0},
\label{eq:<L^m>}
\end{align}
with 
\begin{align}
K_L(\bar\theta) 
= \ln \sum_{N_1,\cdots,N_{M}} P(\vec{N}) e^{\bar\theta L(\vec{N})}
= \ln \langle e^{\bar\theta L(\vec{N})} \rangle .
\label{eq:K_Lln}
\end{align}
From Eq.~(\ref{eq:<L^m>}), the generating function 
$K_L(\bar\theta)$ is expanded as 
\begin{align}
K_L(\bar\theta) = \sum_{m=1}^\infty \frac{\bar\theta^m}{m!} 
\langle (L(\vec{N}))^m \rangle_{\rm c}.
\label{eq:K_Lexp}
\end{align}
Note that the sum starts from $m=1$ because $K_L(0)=0$ which is 
ensured from the fundamental property of probability 
$\sum_{\vec{N}} P(\vec{N})=1$.

By substituting $\bar\theta=1$ 
in Eqs.~(\ref{eq:K_Lln}) and (\ref{eq:K_Lexp}) we obtain,
\begin{align}
\ln \langle e^{L(\vec{N})} \rangle
= \sum_{m=1}^\infty \frac1{m!} 
\langle (L(\vec{N}))^m \rangle_{\rm c}.
\label{eq:cumulantexp}
\end{align}
Equation~(\ref{eq:cumulantexp}) is referred to as 
the cumulant expansion \cite{Asakawa:2015ybt}, 
and plays a crucial role in the following derivations.

\subsection{Binomial distribution function}

The cumulant generating function of the binomial distribution 
function $B_{p,N}(n)$ is given by
\begin{align}
k_N(\theta) = \ln \sum_n e^{\theta n} B_{p,N}(n).
\label{eq:k_N}
\end{align}
By taking derivatives, the $m$-th order cumulant is given by 
$\langle n^m \rangle_{\rm c,binomial} = \xi_m(p) N$ with 
\begin{align}
\xi_m(p) = \frac1N \frac{\partial^m}{\partial \theta^m} k_N(0).
\label{eq:xi(p)}
\end{align}
Explicit forms of $\xi_m(p)$ up to sixth order are given by
\begin{align}
\xi_1(p) =& p,
\label{eq:xi1}
\\
\xi_2(p) =& p(1-p),
\\
\xi_3(p) =& p(1-p)(1-2p),
\\
\xi_4(p) =& p(1-p)(1-6p+6p^2),
\\
\xi_5(p) =& p(1-p)(1-2p)(1-12p+12p^2),
\\
\xi_6(p) =& p(1-p)(1-30p+150p^2-240p^3+120p^4).
\label{eq:xi6}
\end{align}
Using $\xi_m(p)$, Eq.~(\ref{eq:k_N}) is written as 
\begin{align}
k_N(\theta) = \sum_m \frac{\theta^m}{m!} \xi_m(p) N, 
\label{eq:ktheta}
\end{align}
which shows that $k_N(\theta)$ is proportional to $N$.
In this study, we fully make use of this property of $k_N(\theta)$.

\section{Single variable case}
\label{sec:single}

Before the full derivation of Eqs.~(\ref{eq:Q1}) -- (\ref{eq:Q4}),
in this section we first deal with a simplified problem 
with $M=1$, as this analysis would become a good 
exercise for the full derivation addressed in the next section.

In this section, we consider probability distribution functions
$P(N)$ and $\tilde{P}(n)$ for single stochastic variables
$N$ and $n$, respectively, which are related with each other as
\begin{align}
\tilde{P}(n) = \sum_N P(N) B_{p,N}(n).
\label{eq:Psingle}
\end{align}
The cumulant generating function of $\tilde{P}(n)$
is calculated to be
\begin{align}
\tilde{K}(\theta) 
&= \ln \sum_n e^{\theta n} \tilde{P}(n)
\nonumber \\
&= \ln \sum_n e^{\theta n} \sum_N P(N) B_{p,N}(n)
\nonumber \\
&= \ln \sum_N P(N) \sum_n e^{\theta n} B_{p,N}(n)
\nonumber \\
&= \ln \sum_N P(N) e^{k_N(\theta)}
\nonumber \\
&= \ln \langle e^{k_N(\theta)} \rangle,
\label{eq:tildeKsingle}
\end{align}
where in the fourth equality we used Eq.~(\ref{eq:k_N}).
The expectation value in the last line is taken for $P(N)$.

Using the fact that $k_N(\theta)$ is linear in $N$,
Eq.~(\ref{eq:tildeKsingle}) is expressed by the cumulant 
expansion Eq.~(\ref{eq:cumulantexp}) as 
\begin{align}
\tilde{K}(\theta) 
= \sum_m \frac1{m!} \langle (k_N(\theta))^m \rangle_{\rm c},
\label{eq:Kk}
\end{align}
while $\tilde{K}(\theta)$ also defines 
the cumulants of $\tilde{P}(n)$ by their derivatives 
with $\theta=0$,
\begin{align}
\llangle n^m \rrangle_{\rm c} = \partial^n \tilde{K} ,
\label{eq:<n^m>c}
\end{align}
where $\partial=\partial/\partial\theta$ and 
we introduced the following notations:
\begin{enumerate}
\item
The cumulants of single and double brackets are taken 
for $P(\vec{N})$ and $\tilde{P}(\vec{n})$, respectively.
\item
When the argument of generating functions $\tilde{K}(\theta)$ and
$k_N(\theta)$ is suppressed, it is understood that $\theta=0$ is
substituted.
\end{enumerate}
These notations are used throughout this paper.

From Eqs.~(\ref{eq:Kk}) and (\ref{eq:<n^m>c}), 
the $j$-th order cumulant $\llangle n^j \rrangle_{\rm c}$
is represented by the cumulant expansion 
\begin{align}
\llangle n^j \rrangle_{\rm c} 
= \partial^j \sum_m \frac1{m!} \langle k_N^m \rangle_{\rm c}.
\label{eq:<n^m>c=<>}
\end{align}
We now represent the right-hand side of Eq.~(\ref{eq:<n^m>c=<>}) 
by the cumulants of $N$.
To proceed this analysis, there are two convenient rules.
\begin{description}
\item[Rule 1]
$\theta$ derivatives on $\langle k_N^m \rangle_{\rm c}$ act on 
$k_N$'s as if $\langle \cdot \rangle_{\rm c}$ were a standard bracket;
for example,
\begin{align}
\partial \langle k^m \rangle_{\rm c}
=& m \langle k^{m-1} (\partial k) \rangle_{\rm c} .
\\
\partial^2 \langle k^m \rangle_{\rm c}
=& m(m-1) \langle k^{m-2} (\partial k)^2 \rangle_{\rm c}
+ m \langle k^{m-1} (\partial^2 k) \rangle_{\rm c}.
\end{align}
This is because $k_N(\theta)$ is proportional to $N$ and only the 
coefficient in front of $N$ depends on $\theta$.
\item[Rule 2]
By substituting $\theta=0$, 
all $k_N(\theta)$'s which do not receive $\theta$ derivative 
vanish. Therefore, all $k(\theta)$ must receive 
at least one differentiation so that the term gives nonzero 
contribution to $\langle n^j \rangle_{\rm c}$ in Eq.~(\ref{eq:<n^m>c=<>}).
This immediately means that the $m$-th order term in 
Eq.~(\ref{eq:<n^m>c=<>}) can affect $\langle n^j \rangle_{\rm c}$ 
only if $m\le j$.
\end{description}

To see the manipulation for the right-hand side of
Eq.~(\ref{eq:<n^m>c=<>}) with these rules, let us consider the $j=3$ case.
In this case, the terms with $m=1,~2$, and $3$ have nonvanishing
contribution because of Rule 2.
These terms are calculated to be
\begin{align}
\partial^3 \langle k_N \rangle_{\rm c}
=& \langle \partial^3 k_N \rangle_{\rm c}
= \xi_3 \langle N \rangle_{\rm c},
\\
\frac12 \partial^3 \langle k_N^2 \rangle_{\rm c}
=& 3 \langle (\partial^2 k_N)(\partial k_N)  \rangle_{\rm c}
= 3 \xi_2 \xi_1 \langle N^2  \rangle_{\rm c},
\\
\frac1{3!} \langle \partial^3 k_N^3 \rangle_{\rm c}
=& \langle (\partial k_N )^3 \rangle_{\rm c}
= \xi_1^3 \langle N^3 \rangle_{\rm c},
\end{align}
where all terms with $k_N$ without derivatives are neglected
from Rule 2.
Substituting them in Eq.~(\ref{eq:<n^m>c=<>}) we obtain
\begin{align}
\llangle n^3 \rrangle_{\rm c}
=& \partial^3 \tilde{K}
\nonumber \\
=& \partial^3 \langle k_N \rangle_{\rm c}
+ \frac12 \partial^3 \langle k_N^2 \rangle_{\rm c}
+ \frac1{3!} \partial^3 \langle k_N^3 \rangle_{\rm c}
\nonumber \\
=& \xi_3 \langle N \rangle_{\rm c}
+ 3 \xi_2 \xi_1 \langle N^2 \rangle_{\rm c}
+ \xi_1^3 \langle N^3 \rangle_{\rm c}.
\end{align}
Similar manipulations up to $6$-th order lead to 
\begin{widetext}
\begin{align}
\llangle n \rrangle_{\rm c} 
=& \xi_1 \langle N \rangle_{\rm c} ,
\label{eq:<n1>c}
\\
\llangle n^2 \rrangle_{\rm c} 
=& \xi_2 \langle N \rangle_{\rm c} + \xi_1^2 \langle N^2 \rangle_{\rm c} ,
\\
\llangle n^3 \rrangle_{\rm c} 
=& \xi_3 \langle N \rangle_{\rm c} + 3 \xi_2 \xi_1 \langle N^2 \rangle_{\rm c} 
+ \xi_1^3 \langle N^3 \rangle_{\rm c} ,
\\
\llangle n^4 \rrangle_{\rm c} 
=& \xi_4 \langle N \rangle_{\rm c} 
+ (4 \xi_3 \xi_1 + 3 \xi_2^2 ) \langle N^2 \rangle_{\rm c} 
+ 6 \xi_2 \xi_1^2 \langle N^3 \rangle_{\rm c} 
+ \xi_1^4 \langle N^4 \rangle_{\rm c} ,
\\
\llangle n^5 \rrangle_{\rm c} 
=& \xi_5 \langle N \rangle_{\rm c} 
+ (5 \xi_4 \xi_1 + 10 \xi_3 \xi_2 ) \langle N^2 \rangle_{\rm c} 
+ (10 \xi_3 \xi_1^2 + 15 \xi_2^2 \xi_1 ) \langle N^3 \rangle_{\rm c} 
+ 10 \xi_2 \xi_1^3 \langle N^4 \rangle_{\rm c} 
+ \xi_1^5 \langle N^5 \rangle_{\rm c} ,
\\
\llangle n^6 \rrangle_{\rm c} 
=& \xi_6 \langle N \rangle_{\rm c} 
+ (6 \xi_5 \xi_1 + 15 \xi_4 \xi_2 + 10 \xi_3^2 ) \langle N^2 \rangle_{\rm c} 
+ (15 \xi_4 \xi_1^2 + 60 \xi_3 \xi_2 \xi_1 + 15 \xi_2^3 ) \langle N^3 \rangle_{\rm c} ,
\nonumber \\
&
+ (20 \xi_3 \xi_1^3 + 45 \xi_2^2 \xi_1^2 ) \langle N^4 \rangle_{\rm c} 
+ 15 \xi_2 \xi_1^4 \langle N^5 \rangle_{\rm c} 
+ \xi_1^6 \langle N^6 \rangle_{\rm c} .
\label{eq:<n6>c}
\end{align}
\end{widetext}
These are the formulas which represent the cumulants of 
observed particles numbers, $\llangle n^m \rrangle_{\rm c}$, 
using the original ones $\langle N^m \rangle_{\rm c}$.

For the efficiency correction,
we have to represent $\langle N^m \rangle_{\rm c}$
using $\llangle n^m \rrangle_{\rm c}$.
These relations are most straightforwardly obtained by 
representing Eqs.~(\ref{eq:<n1>c}) -- (\ref{eq:<n6>c})
in a matrix form,
\begin{align}
\vec{V}_n =& \mathbb{M} \vec{V}_N,
\label{eq:V_n=MV}
\end{align}
with 
\begin{align}
\vec{V}_n = ( \llangle n \rrangle_{\rm c} , \cdots , \llangle n^m \rrangle_{\rm c} )^T,
\quad
\vec{V}_N = ( \langle N \rangle_{\rm c} , \cdots ,
\langle N^m \rangle_{\rm c} )^T,
\nonumber 
\end{align}
and taking the inverse.
We note that the matrix $\mathbb{M}$ in Eq.~(\ref{eq:V_n=MV}) 
is lower triangular.
Accordingly, the inverse of $\mathbb{M}$ is also lower triangular.
The $m$-th order cumulant $\langle N^m \rangle_{\rm c}$ thus is 
represented 
by $\llangle n^l \rrangle_{\rm c}$ with $l\le m$.
The results correspond to a special case of
Eqs.~(\ref{eq:Q1}) -- (\ref{eq:Q4}) with $M=1$ and $a_1=1$.
The explicit forms up to fourth order are found 
in Ref.~\cite{Asakawa:2015ybt}.

\section{Multi-variable case}
\label{sec:mult}

Next, we derive Eqs.~(\ref{eq:Q1}) -- (\ref{eq:Q4}).
We start from the cumulant generating function of
$\tilde{P}(\vec{n})$ in Eq.~(\ref{eq:tildeP}),
\begin{align}
\tilde{K}(\vec{\theta}) 
=& \sum_{n_1,\cdots,n_M} \tilde{P}(\vec{n}) 
\exp ( \sum_i \theta_i n_i ) 
\nonumber \\
=& \ln \sum_{N_1,\cdots,N_M} P(\vec{N}) 
  \sum_{n_1,\cdots,n_M} \prod_i \big( e^{\theta_i n_i} B_{\epsilon_i,N_i}(n_i) \big)
\nonumber \\
=& \ln \sum_{N_1,\cdots,N_M} P(\vec{N}) 
  \prod_i \big( \sum_{n_i} e^{\theta_i n_i} B_{\epsilon_i,N_i}(n_i)\big) 
\nonumber \\
=& \ln \sum_{N_1,\cdots,N_M} P(\vec{N}) e^{\kappa(\vec{\theta})} 
\nonumber \\
=& \ln \langle e^{\kappa(\vec{\theta})} \rangle_{\rm c},
\end{align}
where
\begin{align}
\kappa(\vec{\theta}) = \sum_{i=1}^M k_{N_i} (\theta_i),
\label{eq:kappa}
\end{align}
with $k_{N_i}(\theta_i)$ being the cumulant generating function 
of $B_{\epsilon_i,N_i}(n_i)$ defined in Eq.~(\ref{eq:k_N}).
In the following, we use the notations 1 and 2 introduced in 
the previous section.

From the definition, it is clear 
that $\kappa$ is a linear function of $N_i$.
Derivatives of $\kappa(\vec{\theta})$ for $\theta_i=0$ are given by
\begin{align}
\partial_i^m \kappa = \xi_m^{(i)} N_i ,
\label{eq:dkappa=xiN}
\end{align}
with $\xi_m^{(i)} = \xi_m(\epsilon_i)$, while 
derivatives of $\kappa(\vec{\theta})$ with 
different $\theta_i$'s vanish, i.e.
\begin{align}
\partial_j^m \partial_k^l \kappa = 0 ,
\label{eq:ddk}
\end{align}
for all $j\ne k$ and nonzero $m$ and $l$, and so forth,
which is trivial from Eq.~(\ref{eq:kappa}).
We also note that  $\kappa(\vec{0})=0$.

Because of the linearity of $\kappa(\vec{\theta})$ on $\vec{N}$,
$\tilde{K}(\vec{\theta})$ can be written in the 
cumulant expansion as 
\begin{align}
\tilde{K}(\vec{\theta}) = \sum_m \frac1{m!} 
\langle (\kappa(\vec{\theta}))^m \rangle_{\rm c} .
\label{eq:Kkappa}
\end{align}
Equations~(\ref{eq:Q1}) -- (\ref{eq:Q4}) are obtained by 
taking appropriate derivatives of Eq.~(\ref{eq:Kkappa})
similarly to the previous section.
In this manipulation, one can again apply the Rules 1 and 2 
introduced in the previous section because of 
the linearity of $\kappa(\vec{\theta})$.

In order to obtain the cumulants of $Q$ defined in Eq.~(\ref{eq:Q}), 
we apply a differential operator,
\begin{align}
D = \sum_i \tilde{a}_i \partial_i ,
\end{align}
to both sides in Eq.~(\ref{eq:Kkappa}),
with $\tilde{a}_i=a_i/\epsilon_i$.
The left-hand side is then given by
\begin{align}
D^m \tilde{K} = \llangle ( \sum_i \tilde{a}_i n_i)^m \rrangle_{\rm c}
= \llangle q_{(1)}^m \rrangle_{\rm c},
\label{eq:DmKq}
\end{align}
where $q_{(1)}$ is defined in Eq.~(\ref{eq:q(s)}).

Next, we see derivatives of the right-hand side order by order.
For the first derivative, only the first term in Eq.~(\ref{eq:Kkappa})
has nonzero contribution and we obtain
\begin{align}
D \tilde{K} = 
\langle D \kappa(\vec{\theta}) \rangle_{\rm c}
= \langle Q \rangle_{\rm c} .
\label{eq:Dkappa}
\end{align}
From Eq.~(\ref{eq:DmKq}) with $m=1$ and Eq.~(\ref{eq:Dkappa}),
we obtain Eq.~(\ref{eq:Q1}).

The second derivative is calculated to be
\begin{align}
D^2\tilde{K} 
= D^2 \big( \langle \kappa \rangle_{\rm c} + 
\frac12 \langle \kappa^2 \rangle_{\rm c} \big)
= \langle D^2 \kappa \rangle_{\rm c} + \langle ( D\kappa )^2 \rangle_{\rm c},
\label{eq:D^2tildeK}
\end{align}
where we used Rules 1 and 2.
The second term in the far right-hand side 
is $\langle Q^2 \rangle_{\rm c}$ as a special case of the relation
\begin{align}
\langle ( D\kappa )^m \rangle_{\rm c} = \langle Q^m \rangle_{\rm c}.
\label{eq:Dkappa^m}
\end{align}
The first term, on the other hand, needs a further manipulation.
For this calculation, we note the relation
\begin{align}
D^m \kappa 
=& \big( \sum_i \tilde{a}_i \partial_i \big)^m \kappa 
= \sum_i \tilde{a}_i^m \partial_i^m \kappa 
= \sum_i \tilde{a}_i^m \xi_m^{(i)} N_i
\nonumber \\
=& \sum_i \tilde{a}_i^m \tilde\xi_m^{(i)} \partial_i \kappa 
= D_{(m)} \kappa,
\label{eq:D2kappa}
\end{align}
where we have defined a differential operator 
\begin{align}
D_{(s)} = \sum_i \tilde{a}^s \tilde\xi_s^{(i)} \partial_i,
\end{align}
with $\tilde\xi_s^{(i)} = \xi_s^{(i)} /\xi_1^{(i)}$.
Note that $D=D_{(1)}$.
In the second equality in Eq.~(\ref{eq:D2kappa}) we have used
Eq.~(\ref{eq:ddk}).
We then take $D_{(2)}$ derivative of $\tilde{K}$ as
\begin{align}
D_{(2)} \tilde{K} = \langle D_{(2)} \kappa \rangle_{\rm c} .
\label{eq:D_2tildeK}
\end{align}
Substituting Eqs.~(\ref{eq:D2kappa}) and (\ref{eq:D_2tildeK}) 
in Eq.~(\ref{eq:D^2tildeK}) and 
\begin{align}
D_{(m)} \tilde{K} = \llangle q_{(m)} \rrangle_{\rm c},
\label{eq:D(m)K}
\end{align}
we obtain Eq.~(\ref{eq:Q2}).

To extend the calculation to third and higher orders,
we need another differential operator
\begin{align}
D_{(s_1,\cdots,s_j|t_1,\cdots,t_k)} 
= \sum_i c_{(s_1,\cdots,s_j|t_1,\cdots,t_k)}^{(i)} \partial_i,
\label{eq:D(sstt)}
\end{align}
which appears in differentiations 
\begin{align}
D_{(s_1)} D_{(s_2)} \kappa =& D_{(s_1,s_2|2)} \kappa ,
\\
D_{(s_1)} D_{(s_2)} D_{(s_3)} \kappa =& D_{(s_1,s_2,s_3|3)} \kappa ,
\end{align}
and
\begin{align}
D_{(\vec{s}_1|\vec{t}_1)} D_{(\vec{s}_2|\vec{t}_2)} \kappa
&= D_{(\vec{s}_1,\vec{s}_2|\vec{t}_1,\vec{t}_2,2)} \kappa,
\\
D_{(\vec{s}_1|\vec{t}_1)} D_{(\vec{s}_2|\vec{t}_2)} D_{(\vec{s}_3|\vec{t}_3)} \kappa
&= D_{(\vec{s}_1,\vec{s}_2,\vec{s}_3|\vec{t}_1,\vec{t}_2,\vec{t}_3,3)} \kappa,
\end{align}
and etc., where the vectorical representations for subscripts are
understood.
On the other hand, the operations of these operators on $\tilde{K}$ give 
\begin{align}
D_{(\vec{s}|\vec{t})} \tilde{K}
&= \llangle q_{(\vec{s}|\vec{t})} \rrangle_{\rm c},
\\
D_{(\vec{s}_1|\vec{t}_1)} D_{(\vec{s}_2|\vec{t}_2)} \tilde{K}
&= \llangle q_{(\vec{s}_1|\vec{t}_1)} q_{(\vec{s}_2|\vec{t}_2)} \rrangle_{\rm c},
\end{align}
and so forth.

For third order, all equations required to obtain 
$\langle Q^3 \rangle_{\rm c}$ are
\begin{align}
D^3 \tilde{K}
=& \llangle q_{(1)}^3 \rrangle_{\rm c}
\nonumber \\
=& \langle D_{(3)} \kappa \rangle_{\rm c}
+ 3 \langle (D_{(2)}\kappa) (D_{(1)}\kappa) \rangle_{\rm c}
\nonumber \\
& + \langle (D_{(1)}\kappa)^3 \rangle_{\rm c},
\label{eq:D^3tildeK1}
\\
D_{(2)} D_{(1)} \tilde{K}
=& \llangle q_{(2)} q_{(1)} \rrangle_{\rm c}
\nonumber \\
=& \langle D_{(2,1|2)}\kappa \rangle_{\rm c}
+ \langle (D_{(2)}\kappa) (D_{(1)}\kappa) \rangle_{\rm c},
\label{eq:D^3tildeK2}
\\
D_{(2,1|2)} \tilde{K}
=& \llangle q_{(2,1|2)} \rrangle_{\rm c}
= \langle D_{(2,1|2)}\kappa \rangle_{\rm c}
\label{eq:D^3tildeK3}
\\
D_{(3)} \tilde{K} =& \llangle q_{(3)} \rrangle_{\rm c}
= 
\langle D_{(3)}\kappa \rangle_{\rm c}.
\label{eq:D^3tildeK4}
\end{align}
Using these results with Eq.~(\ref{eq:Dkappa^m}),
we obtain Eq.~(\ref{eq:Q3}).
Note that the calculation to obtain Eq.~(\ref{eq:Q3})
from Eqs.~(\ref{eq:D^3tildeK1}) -- (\ref{eq:D^3tildeK4})
is straightforwardly carried out by representing 
these equations in a matrix form,
\begin{align}
\llangle V_q  \rrangle_{\rm c} 
= \mathbb{M}_3 \langle V_\kappa \rangle_{\rm c},
\label{eq:M3}
\end{align}
with 
\begin{align}
V_q =& ( q_{(1)}^3 , q_{(2)} q_{(1)} , q_{(2,1|2)} , q_{(3)} )^T,
\\
V_\kappa =& ( (D_{(1)}\kappa)^3 , (D_{(2)}\kappa) (D_{(1)}\kappa) ,
D_{(2,1|2)}\kappa , D_{(3)}\kappa )^T,
\end{align}
and taking the inverse of $\mathbb{M}_3$.

Similarly, the $m$-th order relation for $m\ge4$
is obtained by the following procedures:
\begin{enumerate}
\item
Calculate $D^m \tilde{K}$. 
The result contains $\langle (D_{(1)}\kappa)^m \rangle_{\rm c} 
= \langle Q^m \rangle_{\rm c}$.
\item
The remaining terms in the above result consists of 
derivatives of $\kappa$.
Calculate the derivatives of $\tilde{K}$ with the same differential
operator. For example, when one obtains
$\langle ( D_{(2)}\kappa )( D_{(1)}\kappa )^2 \rangle_{\rm c}$
in the first process, calculate $D_{(2)} D_{(1)}^2 \tilde{K}$.
The result contains the original term (in the example,
$\langle ( D_{(2)}\kappa )( D_{(1)}\kappa )^2 \rangle_{\rm c}$).
\item
Repeat this procedure until all derivatives of $\kappa$ 
are represented by derivatives of $\tilde{K}$.
\item
Unify them in a matrix form like Eq.~(\ref{eq:M3}), 
and take the inverse.
\end{enumerate}
For fourth order, Eq.~(\ref{eq:Q4}) is obtained by 
calculating the following $11$ 
differentiations of $\tilde{K}$:
\begin{align}
& D^4 ,~D_{(3)}D_{(1)},~ D_{(2)}^2,~
D_{(3,1|2)} ,~ D_{(2,2|2)} 
\nonumber \\
& D_{(2)} D_{(1)}^2 ,~ D_{(2,1|2)} D_{(1)} ,~ D_{(1,1|2)} D_{(2)} ,
\nonumber \\
& D_{(2,1,1|2,2)} ,~ D_{(2,1,1|3)} ,~ D_{(4)}.
\end{align}

\section{Discussions}
\label{sec:summary}

In this paper we have presented the formulas representing
the cumulants of $Q$ up to fourth order.
These results can straightforwardly be extended to much higher
orders and to mixed cumulants, though the calculation
becomes more lengthy as the order becomes higher.

In this paper, we considered the binomial model Eq.~(\ref{eq:binomial}).
Only a property of the binomial distribution function $B_{p,N}(n)$
used in the derivations of Eqs.~(\ref{eq:Q1}) -- (\ref{eq:Q4}) is 
the fact that the cumulant generating function $k_N(\theta)$ is
proportional to $N$.
Therefore, $B_{p,N}(n)$ can be replaced by other distribution 
functions satisfying this condition by replacing the values 
of $\xi_i$.


In this paper we discussed the efficiency correction assuming the
binomial model Eq.~(\ref{eq:binomial}).
As discussed already, 
this model can be justified when the efficiencies for 
the individual particles can be regarded independent.
In real detectors, however, efficiencies of individual particles
can be correlated.
The effect of such correlations are discussed recently \cite{efficiency},
and it is suggested that a small correlation can give rise to large
discrepancy of the reconstructed values especially for higher orders.
The efficiency corrected cumulants based on the binomial model 
have to be interpreted with this caveat.

In typical detectors in heavy ion collisions,
neutrons cannot be observed.
Because of this problem, the proton number cumulants are 
experimentally analyzed and compared with 
the theoretical studies on the baryon number cumulants.
As suggested in Refs.~\cite{Kitazawa:2011wh,Kitazawa:2012at},
the reconstruction of the baryon number cumulants is possible 
in the binomial model, since the measurement of protons among 
nucleons can be regarded as the $50\%$ efficiency loss.
In this case, the assumption of the independence of efficiencies
is well justified for high-energy collisions because of the 
isospin randomization \cite{Kitazawa:2011wh,Kitazawa:2012at}.
It is an important subject to analyze the baryon number
cumulants experimentally in this method, and compare 
their values directly with theoretical studies.

In real experiments, particle misidentifications and secondary
particles also affect the event-by-event analysis \cite{Ono:2013rma}.
These effects are another issue which has to be taken care of
besides the problem of the efficiency correction.

In this paper, we derived the relations Eqs.~(\ref{eq:Q1}) -- (\ref{eq:Q4})
which relate the ``original'' and ``observed'' cumulants of
particle numbers in event-by-event analyses.
In these formula, the cumulants of original particle numbers
are represented by the mixed cumulants of observed particles.
The number of cumulants does not depend on the number of
different efficiencies, $M$.
These formulas thus would effectively be applied to practical 
analyses of efficiency correction of the cumulants 
with large $M$.

\section*{Acknowledgment}

The author thanks ShinIchi Esumi for inviting him
to ``The second CiRfSE Workshop'' held at Tsukuba University
on Jan. 18-19, 2016, and for his hospitality during his stay.
The author also acknowledges fruitful discussion during 
this stay with S.~Esumi, Xiaofeng Luo, Hiroshi Masui, 
Toshihiro Nonaka, and Tetsuro Sugiura.
This work is supported in part by JSPS KAKENHI Grant
Number 25800148.

\end{document}